\documentclass[%
  journal=jctcce,%
  manuscript=article,%
  layout=twocolumn,%
]{achemso}
\setkeys{acs}{maxauthors=3, etalmode=truncate}

\usepackage[T1]{fontenc} 
\usepackage{graphicx}
\usepackage{amssymb}
\usepackage{booktabs}
\usepackage{array}
\usepackage{mathtools}
\usepackage[range-phrase=\text{--}, range-units=single]{siunitx}
\DeclareSIUnit\angstrom{\text{Å}}
\usepackage{comment}
\usepackage{upgreek}
\usepackage{amsmath}
\usepackage{amsfonts}
\usepackage{amstext}
\usepackage{mathtools}
\usepackage{amssymb}
\usepackage{siunitx}
\usepackage{color}
\usepackage[normalem]{ulem}
\usepackage{float}
\usepackage{multirow}
\usepackage{graphicx}
\usepackage{subfigure}
\usepackage{caption}
\usepackage{ifthen}
\makeatletter
\newcommand{\isSingleColumn}{\equal{\acs@layout}{traditional}}
\newcommand{\isSuppinfo}{\equal{\acs@manuscript}{suppinfo}}

\ifthenelse{\isSingleColumn}%
    {
        \usepackage{setspace}
        \AtBeginDocument{\singlespacing}

        \renewcommand{\maketitle}{\section*{\@title}}
        \usepackage[fontsize=11pt]{fontsize}
    }{
	    \usepackage[fontsize=9pt]{fontsize}
    }%

\ifthenelse{\isSuppinfo}%
  {%
  	\setlength{\bibsep}{4pt plus 0.3ex}
  	\DeclareCaptionTextFormat{baseline}{\baselineskip 4.5mm #1}
  }{%
    \setlength{\bibsep}{0pt plus 0.3ex}%
    
    \DeclareCaptionTextFormat{baseline}{\baselineskip 9.5pt #1}
  }%

\makeatother

\newcommand{\SMmethods}     {S1}
\newcommand{\SMfeatures}    {S2}
\newcommand{\SMhottraj}     {S3}

\author{Daniel Nagel}
\author{Sofia Sartore}
\author{Gerhard Stock}
\email{stock@physik.uni-freiburg.de}
\affiliation{Biomolecular Dynamics, Institute of Physics,
	University of Freiburg, 79104 Freiburg, Germany.}
\date{\today}

\title[Towards a Benchmark for Markov State Models:  \\ The Folding of HP35]%
{Towards a Benchmark for Markov State Models:  \\ The Folding of HP35}

\begin{document}
%
%

\includegraphics[scale=0.9]{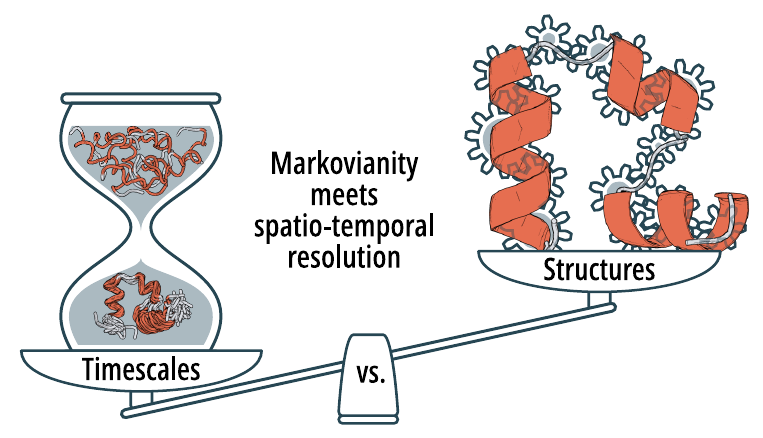}

\begin{abstract}
  Adopting a $300\,\upmu$s-long molecular dynamics (MD) trajectory of
  the reversible folding of villin headpiece (HP35) published by D.\
  E.\ Shaw Research, we recently constructed a Markov state model
  (MSM) of the folding process based on interresidue contacts [J.\
  Chem.\ Theory Comput.\ 2023, {\bf 19}, 3391]. The model reproduces
  the MD folding times of the system and predicts that both the native
  basin and the unfolded region of the free energy landscape are
  partitioned into several metastable substates that are structurally
  well characterized. Recognizing the need to establish
  well-defined but nontrivial benchmark problems, in this Perspective
  we study to what extent and in what sense this MSM may be employed
  as a reference model. To this end, we test the robustness of the MSM
  by comparing it to models that use alternative combinations of
  features, dimensionality reduction methods and clustering schemes.
  The study suggests some main characteristics of the folding of HP35,
  which should be reproduced by any other competitive model of the
  system. Moreover, the discussion reveals which parts of the MSM
  workflow matter most for the considered problem, and illustrates the
  promises and possible pitfalls of state-based models for the
  interpretation of biomolecular simulations.
\end{abstract}

\maketitle
%
%

\section{Introduction}

Molecular dynamics (MD) simulations allow us to study the structure,
dynamics and function of biomolecular systems in atomic
detail.\cite{Berendsen07} To facilitate a concise interpretation of
the resulting 'big data', it is common practice to construct a
coarse-grained model of the considered process, e.g., based on a Langevin
equation\cite{Lange06b, Hegger09, Ayaz21} or a Markov state model
(MSM). \cite{Chodera07,Noe07, Buchete08, Prinz11, Bowman13a,Wang17a,
  Husic18,Noe19} Interpreting MD trajectories in terms of memoryless
transitions between metastable conformational states, MSMs have become
particularly popular as they provide a generally accepted
state-of-the-art analysis of the dynamics,\cite{Bowman13a,Wang17a,
  Husic18} promise to predict long-time dynamics from short
trajectories,\cite{Noe09,Bowman10a} and are straightforward to build
using open-source packages such as PyEmma \cite{Scherer15} and
MSMBuilder.\cite{MSMBuilder} The generally accepted workflow to construct an MSM
consists of five steps: (1) featurization, i.e., selection of suitable
input coordinates, (2) dimensionality reduction from the
high-dimensional feature space to a low-dimensional space of
collective variables, (3) geometrical clustering of these
low-dimensional data into microstates, (4) dynamical clustering of
the microstates into metastable conformational states, and (5)
estimation of the corresponding transition matrix.

In recent years, the groups of No{\'e} and Pande and several others
have established a comprehensive mathematical formulation of Markov
modeling.\cite{Prinz11,Bowman13a,Wang17a, Husic18,Noe19} This includes
their derivation from exact generalized master
equations\cite{Zwanzig83} and transition operator
theory,\cite{Sarich10} their application to nonequilibrium processes,
\cite{Knoch17,Paul19,Hartich21} their combination with multiensemble
\cite{Wu16} and adaptive sampling methods, \cite{Bowman10a} and their
extension to approaches including memory such as core-set MSMs
\cite{Buchete08,Schuette11,Jain14,Lemke16} and memory-enriched
models. \cite{Cao20,Hartich21,Suarez21,Dominic23} In particular, it
has been shown that the quality of an MSM can be optimized by
employing a variational principle,\cite{Nueske14,Wu20} which states
that the MSM producing the slowest implied timescales represents the best
approximation to the true dynamics. As a consequence, the resulting
MSM is dynamically consistent, that is, it reproduces the correct
coarse-grained dynamics, as can be checked by comparison to reference
MD data.
While the variational approach in principle provides a rigorous means
to construct MSMs, its meaningful application to all five steps of the
workflow rests on several assumptions. For example, the considered
process needs to be sufficiently sampled (to ensure that the MD data
are statistically significant) as well as appropriately described by
the chosen input coordinates (to enable a successful model from the
outset).

What is more, we want to assure that the resulting MSM meets its
original purpose, i.e., to explain the mechanism of a biomolecular
process in terms of transitions between functionally relevant and
structurally well-characterized metastable states.  MSMs represent a
spatio-temporal coarse graining of the MD simulation, where the lag
time $\tau_{\rm lag}$ (for which the transition matrix is calculated)
defines the time resolution and the conformational ensembles
represented by the states define the spatial resolution. Hence,
$\tau_{\rm lag}$ needs to be shorter than the fastest dynamics of
interest, and the spatial resolution provided by the state
partitioning needs to be sufficient to account for the relevant
microscopic steps of the functional motion. Unfortunately, these
requests are often in conflict with the requirement of slow implied
timescales enforced by variational methods. For example, while shorter
lag times improve the time resolution, they have the undesirable
effect of shortening the timescales of the MSM, hence resulting in the
loss of Markovianity. Nevertheless, so far most existing strategies to
optimize MSMs (e.g., via the selection of its hyperparameters) are
solely based on the variational approach.\cite{McGibbon15,Husic16}
Hence there is a demand to establish well-defined criteria that ensure
that an MSM also provides the desired microscopic
interpretation. Recently proposed extensions of MSMs including memory
may be a resort in this respect,
\cite{Cao20,Hartich21,Suarez21,Vroylandt22,Dominic23} since they hold
the promise to construct dynamic models with a sufficiently high
spatio-temporal resolution to account in detail for the mechanisms,
while still providing a dynamically consistent description including
the correct long timescales.

At present, the development of post-simulation models such as Langevin
equations and MSMs is at a transition point from gathering concepts
and ideas to a systematic, consistent, and well-understood
methodology. Similar as in more mature research fields such as
electron structure theory and force field development, this requires
an agreement of the community on well-defined benchmark problems, key
observables of interest, and clear criteria of quality
assessment.\cite{note1} This is particularly important, for example,
when we want to assess the rapidly increasing number of
machine-learning empowered methods of conformational analysis (see
Refs.\ \citenum{Wang20,Fleetwood20,Glielmo21,Konovalov21} for
recent reviews).
So far mostly the ``alanine dipeptide'' (sequence Ace-Ala-Nme) has
served as a benchmark problem, because the system is well represented
by two backbone dihedral angles $(\phi,\psi)$ and thus quite simple to
model.\cite{note3,Bolhuis00,Wu22} Moreover various short peptides that
form metastable secondary structures have been considered.

\begin{figure}[ht!]
	\centering
	\includegraphics[scale=0.9]{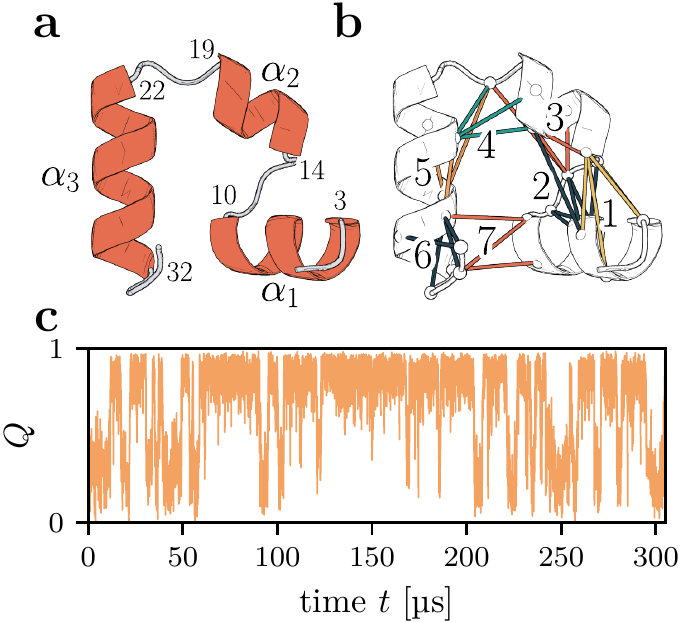}
	\caption{
          The folding of villin headpiece
          (HP35). (a) Molecular structure of the native state,
          consisting of three $\alpha$-helices (residues 3--10, 14--19
          and 22--32) connected by two loops. (b) Illustration of the
          seven main clusters of native contacts, obtained from a
          MoSAIC correlation analysis\cite{Diez22} of the contact
          distances. (c) Time evolution of the fraction of native
          contacts $Q$ obtained from the folding trajectory by Piana
          et al.,\cite{Piana12} showing reversible transitions between
          the unfolded region of the free energy landscape
          ($Q \lesssim 0.4$) and the native basin ($Q \gtrsim 0.7$).}
	\label{fig:HP35}
\end{figure}

To move towards more biophysically relevant systems, the obvious next
step is to establish a benchmark model of a simple globular protein
with a tertiary structure.
As a prime example, here we consider the folding of the 35-residue
villin headpiece, aka HP35, consisting of a hydrophobic core with
three helices that are connected via two short loops (Fig.\
\ref{fig:HP35}a). Specifically, we adopt a $300\,\upmu$s-long
trajectory at 360\,K of the fast folding variant Lys24-Nle/Lys29-Nle,
which is publicly available from D.\ E.\ Shaw Research.\cite{Piana12}
Showing 33 folding and 32 unfolding events (Fig.\ \ref{fig:HP35}c),
the simulation reproduces at least qualitatively some of the main
experimental findings for the system,\cite{McKnight97,Kubelka03,Kubelka04,
  Kubelka06} including the melting temperature (370 K in MD vs.\ 361 K
in experiment), the folding enthalpy (21 vs.\ 25 kcal/mol), as well as
the folding time (1.8 vs.\ 0.7 $\mu$s). Being one of the smallest
naturally occurring proteins that folds autonomously into a globular
structure at record speed,\cite{Kubelka04,Kubelka06} HP35 is probably
the most intensively studied ultrafast folder by experiment,
\cite{Kubelka03,Kubelka04,Kubelka06,Brewer07,
  Kubelka08,Reiner10,Chung11,Serrano12,Eaton21} MD simulation,
\cite{Duan98,Snow02, Fernandez03,Lei07a, Ensign07, Rajan10,Shaw10,
  Piana12} and conformational analysis, \cite{Beauchamp12,Sormani20,
  Chong21,Jain14, Ernst15,Sittel18,Nagel19,Nagel20} and was featured
as 'the new benchmark system of protein folding' by Bill
Eaton.\cite{Eaton21}

While a large number of theoretical analyses have been performed on
the long folding MD trajectories of D.\ E.\ Shaw Research, the
resulting models are often hard to compare and their quality is
difficult to assess. To a large extent this is caused by the very
first step of the workflow, the choice of input coordinates or
features. Since 'you get what you put in', feature selection is
crucial for every conformational analysis.
\cite{Husic16,Sittel18,Scherer19,Ravindra20} Adopting the above
mentioned trajectory of HP35 by Piana et al.,\cite{Piana12} we
recently performed a careful study of the effects of feature selection
on Markov modeling.\cite{Nagel23} As a main result, we showed that
backbone dihedral angles account accurately for the structure of the
native energy basin of HP35, while the unfolded region of the free
energy landscape and the folding process are best described by
tertiary contacts of the protein. In particular, we showed that the
contact-based MSM describes consistently the hierarchical structure of
the free energy landscape, that is, predicts that both the native
basin and the unfolded region are structured into several metastable
substates, evidence of which was shown in various
experiments.\cite{Kubelka06,Brewer07,
  Kubelka08,Reiner10,Chung11,Serrano12,Eaton21}

In this perspective, we wish to investigate to what extent and in what
sense the contact-based MSM of Ref.\ \citenum{Nagel23} might be
established as a reference or benchmark model of the coarse-grained
folding dynamics of HP35. For one, the model satisfies our above
discussed criteria, that is, provides structurally well-characterized
metastable states that account for the folding pathways of HP35 (thus
explaining the underlying mechanism) and whose dynamics compare well
to reference calculations obtained from the MD trajectory (proving the
dynamical consistence of the model). However, it is (most likely) not
the best possible model and certainly not unique, because the MSM
workflow necessitates the choice of numerous methods and metaparameters.
As a minimum requirement for a reference model, we therefore evaluate
the model's robustness with respect to such variations by employing
alternative methods for all steps of the MSM workflow. While it is
neither possible nor desirable to exhaustively sample the
high-dimensional 'method space', the MSMs resulting from a few popular
method combinations nevertheless give a qualitative idea of the
accuracy and the quality that can be achieved by this type of
modeling. In this way, we establish some main characteristics that we
believe are intrinsic to the folding of HP35 (or at least to this
specific trajectory), and therefore should be reproduced by any other
competitive model of the system.

%
%
\section{Results}
\subsection{Reference model} \label{sec:Reference}

\begin{figure*}
	\centering
	\includegraphics[scale=0.7]{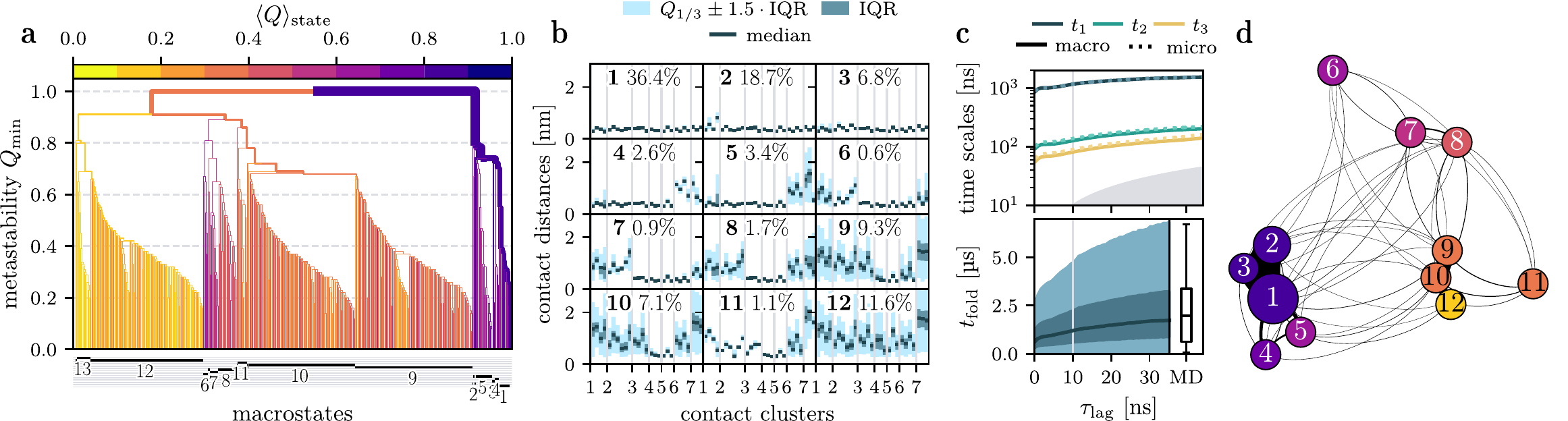}
	\caption{
          Contact-based model of HP35, using
          principal component analysis (PCA), robust density-based
          clustering (RDC) and the most probable path algorithm
          (MPP). (a) MPP dendrogram illustrating the clustering of
          microstates into metastable states upon increasing the
          requested minimum metastability criterion $Q_\text{min}$ of
          a state.  (b) Structural characterization of the twelve
          metastable states of HP35. The states are ordered by
          decreasing fraction of native contacts $Q$, the contacts are
          ordered according to the seven main clusters
          (Fig.~\ref{fig:HP35}b). (c) First three implied timescales
          $t_n$ shown as a function of the lag time
          $\tau_{\text{lag}}$ and box-plot representation of the
          folding time distributions obtained from MSM and MD data.
          (d) Kinetic network of the MSM generated from the
          force-directed algorithm ForceAtlas2.\cite{ForceAtlas2} The
          node size indicates the population of the state and the
          color code reflects its fraction of native contacts $Q$.}
	\label{fig:Reference}
\end{figure*}

To define the contact-based MSM of HP35, we briefly introduce the
methods used in the MSM workflow (for further details see Ref.\
\citenum{Nagel23}). We begin with the choice of features, where we
focused on native interresidue contacts, which are believed to largely
determine the folding pathways of a protein,
\cite{Sali94,Wolynes95,Best13} as they account for the mechanistic
origin of the studied process. Moreover, the fraction of native
contacts represents a well-established one-dimensional reaction
coordinate,\cite{Best13,Best10} which nicely illustrates the overall
time evolution of the folding (Fig.~\ref{fig:HP35}c). Here we assume a
contact to be formed if the distance $d_{ij}$ between the closest
non-hydrogen atoms of residues $i$ and $j$ is shorter than
\SI{4.5}{\angstrom}, where $d_{ij}$ is the minimal distance between
all atoms pairs of the two residues.\cite{Nagel23,Yao19} To focus on native
contacts, we request that contacts between these atoms pairs are
populated more than \SI{30}{\%} of the simulation time, which excludes
nonnative contacts that are typically infrequent and
short-lived.\cite{Nagel23}
To characterize the resulting 42 native contacts of HP35, we calculate
the linear correlation matrix $\rho$ of their distances, and
block-diagonalize $\rho$ using the Python package
MoSAIC\cite{Diez22} (see Methods). This results in seven main clusters
of highly correlated contacts, which change in a cooperative manner
during the folding process. As the contacts of the clusters follow
nicely the protein backbone from the N- to the C-terminus
(Fig.~\ref{fig:HP35}b), the clusters can be employed to characterize
the structure of the conformational states of the protein (see Fig.\
\ref{fig:Reference} below).

To construct metastable conformational states from the above defined
trajectory of contact distances, the following protocol is used. In a
first step, we eliminate high-frequency fluctuations of the distance
trajectory, by employing a Gaussian low-pass filter with a standard
deviation of $\sigma=2\,$ns. Aiming to avoid the misclassification of
the data points in the transition regions, this simple procedure
(prior to clustering) was found to perform better than using dynamic
core sets \cite{Jain14,Nagel19} following clustering.
For dimensionality reduction we use principal component analysis (PCA)
on the smoothed contact distances. \cite{Ernst15} The first five
components exhibit a multimodal structure of their free energy curves,
reveal the slowest timescales ($\sim 0.1$ to $\SI{2}{\micro\second}$),
and explain $\sim \SI{80}{\%}$ of the total correlation.
Using these collective variables, we next perform robust density-based
clustering\cite{Sittel16} (RDC), which computes a local free energy
estimate for every frame of the trajectory by counting all other
structures inside a hypersphere of fixed radius. By reordering all
structures from low to high free energy, RDC directly yields the
minima of the free energy landscape. Requesting a minimal population
($P_\text{min}=\SI{0.01}{\%}\,\widehat{=}\,\SI{153}{frames}$) each
state must contain, we obtain 547 microstates.

In the next step, we adopt the most probable path
algorithm\cite{Jain12} (MPP) to construct a small number of
macrostates. Starting with the above defined microstates, MPP first
calculates the transition matrix of these states, using a lag time
$\tau_\text{MPP}=\SI{10}{\nano\second}$.  If the self-transition
probability of a given state is lower than a certain metastability
criterion $Q_\text{min} \in (0, 1]$, the state will be lumped with the
state to which the transition probability is the highest.  Repeating
the procedure for increasing $Q_\text{min}$, we construct a dendrogram
that reveals how the various metastable states merge into energy
basins, see Fig.\ \ref{fig:Reference}a.
For $Q_\text{min} \gtrsim 0.9$, we obtain only two macrostates, as all
microstates are assigned to either the native or the unfolded energy
basin of HP35. Coloring the states according to their mean number of
native contacts, the native states are drawn in purple and the
unfolded in yellow to orange. By decreasing the requested
metastability $Q_\text{min}$, we in effect decrease the requested
minimum barrier height between separated states, such that the two
main basins split up in an increasing number of substates. To resolve
at least the first tier of the emerging hierarchical structure, we
require that the states should have at least $Q_\text{min}=0.5$ and a
minimum population of \SI{0.5}{\%}, thus obtaining 12 metastable
states.\cite{note2} Remarkably, we find that the native basin as well as the
unfolded basin splits up in various well-characterized states of high
metastability.

As stressed in the Introduction, the metastable states of an MSM should
represent distinct conformational ensembles, in order to allow for a
detailed characterization of the folding mechanism. To provide such a
structural characterization, Fig.~\ref{fig:Reference}b shows the
distribution of contact distances for each state. The states are
ordered by decreasing fraction of native contacts, such that state\,1
is the native state (all contact distances are shorter than
$\SI{4.5}{\angstrom}$) and state\,12 is the completely unfolded state
with a broad distribution of large distances. The contacts are ordered
according to the seven main MoSAIC clusters shown in
Fig.~\ref{fig:HP35}b, which follow the protein backbone from the N- to
the C-terminus.  The first three states are structurally well-defined
native-like states that differ in details of helix\,1 and contain
\SI{62}{\%} of the total population. From the MPP dendrogram we learn
that states 4 and 5 also belong to the native energy basin, and differ
from state\,1 by broken contacts on the C-terminal side. The unfolded
basin consists of states 9 to 12, which show increasing degree
of disorder. Furthermore, there are three lowly populated
($\lesssim \SI{1}{\%}$) intermediate states.
We note in passing that it is difficult to discriminate the metastable
states via a visual inspection of their three-dimensional structural
ensembles. This is because the main three native states exhibit only
small structural differences, and because the structural ensembles of
the unfolded states show a large variance. On the other hand, the
contact representations in Fig.~\ref{fig:Reference}b provide a concise
structural description of the states, because they directly account
for the most important structural determinants of the folding process.

To assess the dynamical properties of the states, we calculate the
transition matrix describing the probability of a transition between
any two states during some chosen lag time $\tau_{\text{lag}}$.  By
diagonalizing this matrix, we obtain its eigenvalues $\lambda_n$ and
the implied timescales $t_{n} = - \tau_{\text{lag}}/\ln{\lambda_n}$.
To optimally project the microstate dynamics onto the macrostate
dynamics, we use for the calculation of the transition matrix the
method of Hummer and Szabo,\cite{Hummer15} which ensures that the
macrostate timescales closely approximate the microstate timescales.
Figure \ref{fig:Reference}c shows the resulting first three micro- and
macrostate timescales $t_1$--$t_3$, which both are found to level off
for lag times $\tau_{\text{lag}} \gtrsim \SI{10}{\nano\second}$.
For further reference, we summarize the quality of the implied
timescales by a single number, the
generalized matrix Rayleigh quotient (GMRQ),\cite{McGibbon15}
given by the sum of the first $3$ eigenvalues $\lambda_n$.
For $\tau_{\text{lag}} = \SI{10}{\nano\second}$, we obtain
GMRQ$=2.80$. (Note that its theoretical maximum value is 3.)
To see how the implied timescales translate to measurable observables
of the folding process, we also consider the distribution of the
folding time $t_{\text{fold}}$, defined as the waiting time for the
transition of the completely unfolded state\,12 to the native
state\,1. Shown in Fig.~\ref{fig:Reference}c, the resulting folding
time is found to increase only little with $\tau_{\text{lag}}$. While
the median of the MSM prediction of $t_{\text{fold}}$ somewhat
underestimates the MD result, overall the MSM reproduces the rather
broad MD folding-time distributions convincingly.

Let us finally illustrate the MSM with a network, where
the nodes correspond to the states $i$ with equilibrium population
$\pi_i$ and the edge weights $f_{ij}$ to the transition probabilities
$T_{ij}$ between two states (Fig.\ \ref{fig:Reference}d). To define a
kinetic distance between each pair of states,\cite{Noe15} we use the
symmetric edge weight $f_{ij} = \pi_i T_{ij} = \pi_j T_{ji} = f_{ji}$,
which exploits the detailed balance between the two
states.\cite{Nagel20} As anticipated from the MPP dendrogram, the
resulting kinetic network exhibits a native basin comprising states 1
to 5, which is clearly separated from the unfolded basin comprising
states 9 to 12. The closeness of the states within a basin indicates
fast interconversion between these states.  The lowly populated states
6 to 8 are either temporarily visited from the unfolded basin or used
as on-route intermediate states on the folding pathway.

Running a long ($10^9$ steps) Markov chain Monte Carlo simulation of
the MSM, we finally determine the overall folding
mechanism. Starting in the completely unfolded state\,12, the first
step (within 150 ns) is the hydrophobic collapse of the protein, which
mostly leads to the states\,9 and 10 of the unfolded basin, where at
least the contacts connecting helices\,2 and 3 are formed. Taking
about $\SI{1.7}{\micro\second}$, the subsequent escape from the
unfolded basin clearly represents the slowest step of the folding
process, and explains the broad distribution of folding times
(Fig.~\ref{fig:Reference}c). Although this cooperative process may include one or
several of the lowly populated partially unfolded states 4 - 8, the
overall folding transition is cooperative, that is, the majority of
contacts are formed simultaneously (i.e., within nanoseconds).  With
the advent in the native basin, the protein finally relaxes within
$\lesssim 100\,$ns into the native state\,1. The resulting most
frequented folding paths of the MSM consist of permutations of the
states in the native and the unfolded basins as well as of the
intermediate states, and agree well with the folding paths obtained
from the MD trajectory. The structural rearrangement in the unfolded
and native basin is believed to show up as a fast ($\sim 100\,$ns)
transient in several experiments.\cite{Kubelka06,Brewer07,
  Kubelka08,Reiner10,Chung11,Serrano12,Eaton21}

We are now in a position to summarize the key findings, which we
believe to be intrinsic to the considered folding trajectory of HP35.
\begin{enumerate}
\item The model exhibits a hierarchical structure of the free energy
  landscape (Fig.~\ref{fig:Reference}a), where both the native and the
  unfolded basins split up in various well populated metastable
  states. Moreover, a few lowly populated intermediate states exist.

\item The contact representation of these states in
  Fig.~\ref{fig:Reference}b reveals that the metastable states are
  structurally well characterized and distinct.

\item The MSM reproduces the slow timescales of the folding process
  (Fig.~\ref{fig:Reference}c). 

\item The folding starts with the hydrophobic collapse to pre-folded
  states in the unfolded basin (150 ns), the escape from which
  represents the slowest step ($\SI{1.7}{\micro\second}$) of the
  process, and ends with the structural relaxation (100 ns) in the
  native basin. The folding transition is cooperative, and the main
  pathways of the MSM agree well with the folding paths of the MD
  trajectory.
\end{enumerate}

As discussed in the Introduction, the performance of an MSM
may crucially depend on small details of the used methods and the
associated metaparameters. To establish the contact-based MSM as a
benchmark system for the folding of HP35, we therefore need to test the
robustness of the model, that is, we validate the results above by
employing alternative methods for the construction of the MSM.

%
%
\subsection{Effect of dimensionality reduction and clustering methods}

\begin{figure}[ht!]
	\centering
	\includegraphics[scale=0.8]{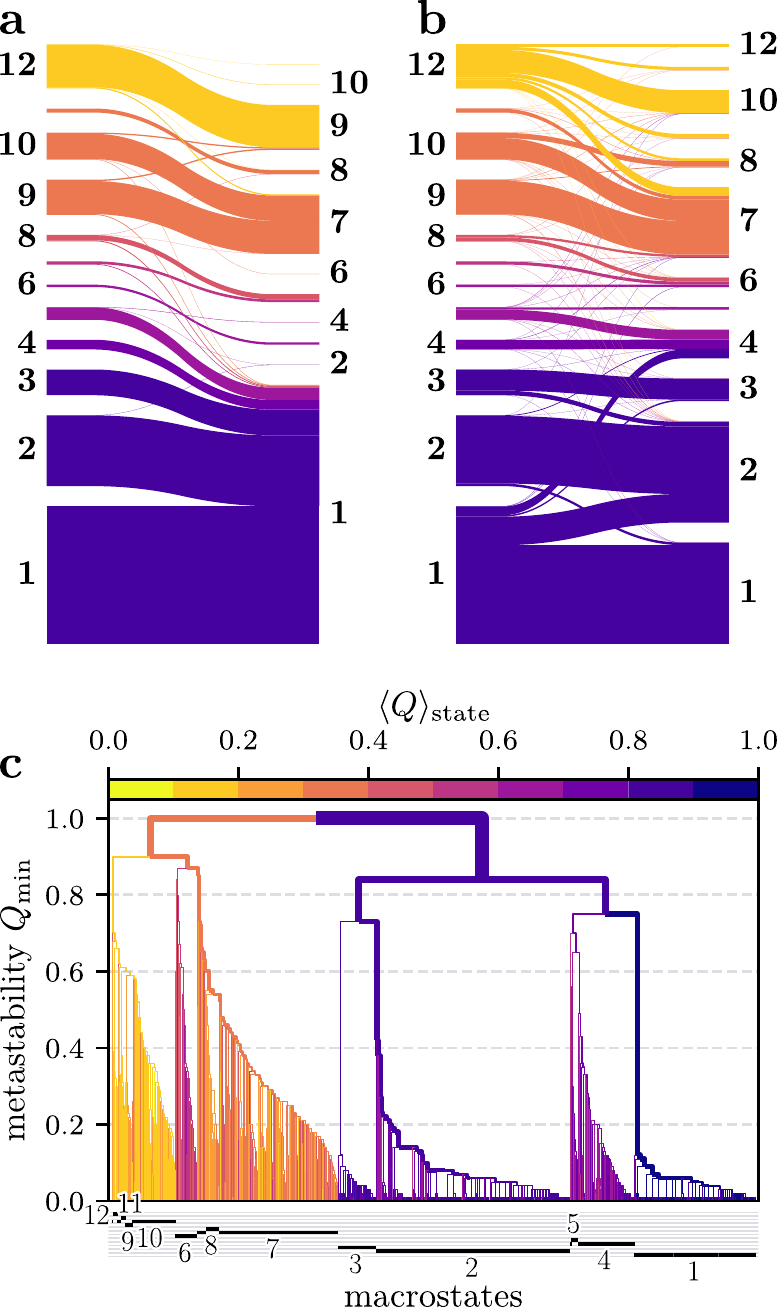}
	\caption{
          (a) Effects of using G-PCCA
          instead of MPP for the construction of macrostates. The
          Sankey diagrams compare the states obtained for the
          reference model (PCA/RDC/MPP, on the left) and the
          alternative model (PCA/RDC/G-PCCA, on the
          right). (b) Performance of the method combination
          tICA/$k$-means/MPP as revealed by the Sankey plot
          comparison to the reference model and (c) the MPP dendrogram.}
	\label{fig:workflow}
\end{figure}

While a large number of techniques for dimensionality reduction,
\cite{Rohrdanz13,Fiorin13,Rodriguez18,Chen18,Lemke19,Wang19,Wang20,
  Fleetwood20,Glielmo21,Konovalov21} and the construction
of microstates \cite{Keller10,Rodriguez14,Song17,Westerlund19} and
macrostates \cite{Bowman13,Roeblitz13,Reuter19,Martini17,Wang18}
exist, here we focus on the most wildly used methods, that is,
time-lagged independent component analysis \cite{Perez-Hernandez13}
(tICA) as an alternative to PCA, $k$-means clustering
\cite{Sculley10,Arthur07,Pedregosa11} as an alternative to RDC, and
the generalized Perron cluster cluster analysis\cite{Reuter19}
(G-PCCA) as an alternative to MPP (see Methods for details).
For the resulting eight combinations of methods, we show in Fig.\
\SMmethods\ the MPP dendrogram, the contact characterization of the
metastable states, a Sankey diagram to compare to the reference
states, and the implied timescales. To summarize the main findings, we
first consider a few key results comprised in Fig.\ \ref{fig:workflow}.

It is instructive to begin at the end of the MSM workflow and consider
the G-PCCA algorithm as an alternative for the generation of
macrostates. Performing spectral clustering of the microstate
transition matrix, the approach aims to maximize the implied
timescales of the resulting macrostates by focusing on the largest
eigenvalues. The optimal number of macrostates, $m$, can be estimated
from the crispness parameter, \cite{Roeblitz13} which suggests that
$m \!=\! 2$ is the optimal choice for the given microstate trajectory,
followed by $m \!= \!3, 4, 5, \ldots$.
Comparing the resulting G-PCCA macrostates to our reference model, we
find for $m = 2$ that G-PCCA partitions all native states (states 1–5
obtained by the MPP algorithm) into a single state and all remaining
ones into the other. Starting with $m \!=\! 3$, the MPP states 6–8 are
assigned as an own state, and from $m \!=\! 4$ on the denatured basin
is divided into a completely unfolded state (state 12) and a partially
unfolded state (states 9–11).

Using larger values for $m$, however, we do not achieve the expected
further fine-graining of the state space, because all additionally
generated states are found to be hardly populated ($\lesssim 1\,\%$).
This is revealed by the Sankey plot in Fig.~\ref{fig:workflow}a, which
compares the 12-state reference model to the G-PCCA results. Choosing
$m \!=\! 11$ for best comparison, we find that the four main states
(already obtained for $m \!=\! 4$) contain 98.0\,\% of the population. That
is, G-PCCA predicts a single highly populated (68.0\,\%) state for the
native basin (instead of a partitioning into five distinct states),
two (instead of three) lowly populated intermediate states, and three
(instead of four) unfolded states.
Interestingly, we find that the implied timescales associated with
the macrostates obtained from G-PCCA and MPP are virtually the same in
all cases (Fig.\ \SMmethods). That is, for $\tau_{\text{lag}}= 10\,$ns
the overall timescale score GMRQ\cite{McGibbon15} is 2.80 for both
G-PCCA and MPP.

As the above findings could be a result of the special choice of
methods, we also employed G-PCCA together with the method combinations
PCA/$k$-means, tICA/RDC, and tICA/$k$-means (Fig.\
\SMmethods{c,d}). Using the popular combination tICA/$k$-means, for
example, we find a quite similar state partitioning as discussed above
for PCA/RDC, i.e., only a single native state, one lowly populated
intermediate state, and three unfolded states. While the results for
tICA/RDC are again comparable, the combination PCA/$k$-means
appears to be the only one that achieves a splitting of the native
basin into four states, albeit with a different partitioning.
From the above results we conclude that G-PCCA tends to
achieve less spatial resolution than MPP, while the implied
timescales are quite similar.

Going back to using MPP for the construction of macrostates, we next
study the performance of the microstate clustering by
$k$-means (instead of RDC) in some more detail. Considering again the
popular combination tICA/$k$-means, Fig.~\ref{fig:workflow}c shows the
resulting MPP dendrogram, which reveals how the microstates are merged
to macrostates. As expected from a purely geometrical clustering
scheme, the $k\!=\!1000$ microstates generated by $k$-means (shown for
$Q_{\rm min}\!=\!0$) typically exhibit significantly lower metastability
than the density-based RDC microstates (Fig.~\ref{fig:Reference}a),
and therefore are lumped already for low values of $Q_{\rm
  min}$. Moreover, $k$-means assigns a similar number of microstates
to the native and the unfolded basins, while the free energy-based RDC
exhibits only a few low-energy states in the native basin.

Comparing the resulting state partitioning to the reference,
the Sankey diagram in Fig.~\ref{fig:workflow}d reveals that the
combination tICA/$k$-means gives a closely related state description of
the native basin, except for some reshuffling between states\,1 and
2. On the other hand, we obtain only a single intermediate state and
four unfolded states that are partitioned in a somewhat different
way. This overall picture is similar for the combination tICA/RDC,
while the combination PCA/$k$-means produces only 8 (instead of 12)
metastable states (Fig.\ \SMmethods). With a GMRQ score of 2.83, 2.80,
and 2.80 for tICA/$k$-means, tICA/RDC, and PCA/$k$-means,
respectively, the implied timescales of all combinations are again
very similar.
Hence we find that the replacement of PCA by tICA and RDC by
$k$-means overall yields similar dynamical models, although with
possibly dissimilar spatial resolution of the free energy landscape. 
As a note of caution, however, we wish to stress that this conclusion
rests heavily on the specific implementation of all used methods as
well as on the choice of metaparameters
(see Methods). Even more so, it is only valid if suitable features are
chosen, as will be shown in the following.

%
%
\subsection{Effect of feature selection}

While the reference model employed native contacts determined via
minimal distances, in the following we discuss several other popular
choices, including Cartesian atom coordinates, C$_\alpha$-distances,
backbone dihedral angles, and a mixture of contacts and angles, see
Fig.\ \SMfeatures. If not noted otherwise, we use our standard
workflow PCA/RDC/MPP and compare it to the popular combination
tICA/$k$-means/MPP.

\begin{figure*}
	\centering
	\includegraphics[scale=0.7]{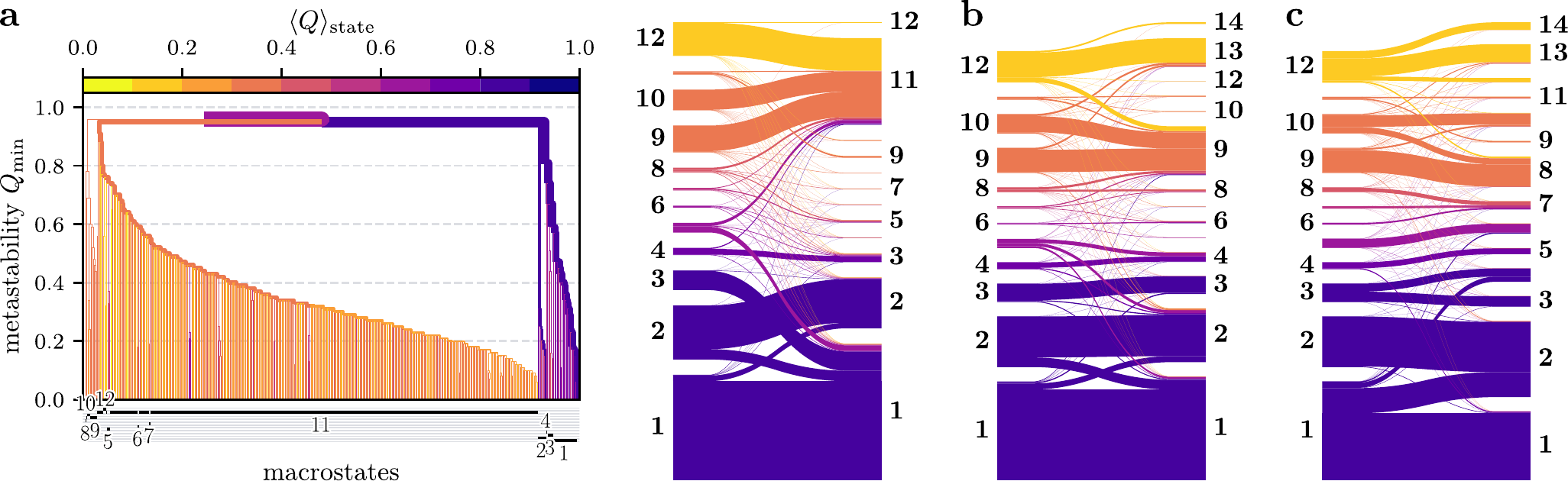}
	\caption{
          Effects of the choice of features,
          including (a) Cartesian coordinates, (b) selected
          C$_\alpha$-distances, and (c) contacts combined with the
          backbone dihedral angle $\phi_3$. Shown are the Sankey plots
          for comparison to the reference states and the MPP
          dendrogram in case (a).}
	\label{fig:features}
\end{figure*}

\subsubsection*{Cartesian atom coordinates}

We first consider Cartesian coordinates of
the C$_\alpha$-atoms, which seem convenient as they directly reveal
the three-dimensional structure of the system. To remove the overall
translation and rotation of the trajectory, the coordinates
of the rigid secondary structures (i.e., the three $\alpha$-helices of
HP35) are typically aligned to a reference structure. As is well
known,\cite{Mu05,Sittel14, Sittel18} however, this procedure cannot
entirely remove the overall rotation of a flexible system. Moreover,
it has been shown that the calculation of the linear correlation
underlying PCA and tICA is ill-defined for three-dimensional vectors,
because the contributions in $x$, $y$ and $z$ direction depend
spuriously on the relative orientation of the
fluctuations.\cite{Lange06}

Displaying the resulting MPP dendrogram and the Sankey plot for
comparison to the reference states, Fig.\ \ref{fig:features}a reveals
the consequences of these issues. Most obviously, the MPP dendrogram
shows that Cartesian coordinates cannot capture the hierarchical
structure of the unfolded part of the free energy landscape, i.e., the
unfolded basin is virtually structureless and represented by only a
single state. Moreover, while the native basin is found to split up in
several states, their contact representation reveals that the
structures of these states are mixed instead of being well
characterized, see Fig.\ \SMfeatures{a} for further details. Reflected
by a GMRQ score of 2.72, the first implied timescale of the model is
too low by a factor 5, which adds to the conclusion that
Cartesian-based MSMs are not suited to describe protein
folding.\cite{Sittel18}
Interestingly, we find that the timescales improve considerably (GMRQ
= 2.90) when we use tICA instead of PCA. Since tICA cannot correct for
the mixing of overall and internal motion, however, the resulting
state partitioning is still rather poor (Fig.\ \SMfeatures{a}). Hence,
we add as a note of caution that variational feature selection can
rate Cartesian-based MSMs deceivingly high,\cite{Scherer19} although
they may not correctly account for the structural evolution during
folding.

%
%
\subsubsection*{C$_\alpha$-distances}

To avoid problems associated with the mixing of overall and internal
motion, internal coordinates are commonly chosen as features. An
obvious option are interresidue distances, such as
C$_\alpha$-distances and contact distances. While the appropriate
identification of contacts via minimal atom distances requires some
effort (see Sec.\ \ref{sec:Reference}), distances between C$_\alpha$
atoms are readily extracted from an MD trajectory and may be used as
an approximate definition of an interresidue contact. This is achieved
by employing a distance cutoff of 8\,\AA, \cite{Yao19} and by
requesting a minimum persistence probability of 30\,\% in order to
focus on native contacts (see Sec.\ \ref{sec:Reference}).

Showing the resulting MPP dendrogram and the Sankey plot, Fig.\
\ref{fig:features}b reveals that the C$_\alpha$-distance-based model
reproduces faithfully the states of the native basin, while the
unfolded part of the energy landscape matches only roughly the
reference results. In particular, we find that the two main unfolded
states are structurally not well defined (Fig.\ \SMfeatures{b}),
whereas the partially and completely unfolded states of the reference
model clearly differ by the existence of contacts in clusters\,3 and 4
(Fig.\ \ref{fig:Reference}b).  This is because the approximate contact
definition is sufficient to distinguish well-defined native states,
but is not suited to partition the structurally very heterogeneous
unfolded basin. As a consequence, we find that the implied timescales
are significantly reduced, yielding a GMRQ score of 2.67. While the
latter issue can be improved by using tICA instead of PCA (resulting
in GMRQ = 2.89), tICA does not yield a better state partitioning of
the unfolded basin.

Besides using selected C$_\alpha$-distances to approximate
interresidue contacts, we can also simply employ all $N(N\!-\!1)/2$
C$_\alpha$-distances as features for an MSM. Although this results in
a highly redundant feature set whose dimension increases rapidly for
larger systems, the choice is quite popular as it is straightforward
and requires no metaparameters.\cite{Scherer19} Somewhat surprisingly,
though, the resulting MPP dendrogram and the Sankey plot in Fig.\
\SMfeatures{c} reveals that the state partitioning deteriorates
dramatically when we use all (instead of selected)
C$_\alpha$-distances. Similar as found above for Cartesian atom
coordinates, the unfolded basin is virtually structureless and
represented by only a single state, while the native basin consists of
three states, whose structures are mixed.  A MoSAIC analysis reveals
the existence of a number of clusters, whose many redundant distances
and the associated ambiguity hamper an accurate structural
characterization. Moreover, the first implied timescale of the model
is too low by a factor 5, giving a poor GMRQ score of 2.47.  While
timescales --and to some extent also the state splitting-- can be
improved by using the combination tICA/$k$-means instead of PCA/RDC,
the various states are structurally still poorly defined (Fig.\
\SMfeatures{c}).

%
%
\subsubsection*{Backbone dihedral angles}

In Ref.\ \citenum{Nagel23} we compared in detail the virtues and
shortcomings of using backbone dihedral angles instead of interresidue
contacts. While dihedral angles are readily obtained from the MD
trajectory, they require an appropriate treatment of their periodicity
\cite{Altis07,Sittel17,Zoubouloglou22} (e.g., by using maximal-gap
shifted\cite{Sittel17} ($\phi, \psi$) dihedral angles) and necessitate
the exclusion of uncorrelated motion of the terminal ends. With
$\psi$-angles reflecting the helicity of the protein and $\phi$-angles
accounting for potential left-to-right handed transitions (e.g., in
flexible loops), backbone dihedral angles report directly on the local
secondary structure. This proves advantageous for the modeling of the
conformational states in the native basin of HP35, which interestingly
can be well approximated by a single angle, $\phi_3$. However, dihedral
angles account only indirectly for the formation of tertiary structure
during folding, which yields a low-resolution description of the
unfolded basin, hampers the precise modeling of the folding
transition, and therefore results in shorter implied
timescales.\cite{Nagel23} 

While the discussion above indicates that interresidue contacts
are better suited for the description of folding, it also suggests
combining the best of two worlds and use contacts to describe the
formation of tertiary structure, and include the single angle $\phi_3$
to accurately account for the substates of the native
basin. (We note in passing that various types of feature
can be combined in the analysis, because we normalize all features
before dimensionality reduction.) Regarding the state partitioning of
the native basin, the resulting Sankey plot is found to differ only
minor from contacts-only reference model (Fig.\ \ref{fig:features}c),
i.e., there is no benefit here. Interestingly, though, the unfolded
part of the free energy landscape is described with high resolution
and shows numerous substates, which might represent an
improvement upon the reference model (Fig.\ \SMfeatures{d}).

%
%
\subsection{Effect of temperature}

An alternative way to assess the robustness of a model is studying to
what extent it accounts for a variation of the input data. As an
example, we consider a folding trajectory of HP35 at 370~K (instead of
360 K), which was also obtained by Piana et al.\cite{Piana12} Since they
otherwise applied identical simulation conditions, we use the same
42 minimal-distance contacts as features.
Showing the time evolution of the fraction of native contacts, Fig.\
\SMhottraj{a} reveals that due to the increased temperature only about
half of the trajectory (instead of two thirds) populates the native
basin, and that the mean folding time is reduced by 20\%. While these
results are expected, it is less clear how the free energy landscape
and the state partitioning changes.
Using our standard workflow PCA/RDC/MPP, the 370~K trajectory results
in virtually the same five native states, while the unfolded part of
the free energy landscape collapses into a single unfolded state
(Fig.\ \SMhottraj{b}). Likewise, the combination tICA/$k$-means/MPP
splits up the unfolded basin in four states (similar to the ones found
for 360~K) and also yields a single unfolded state (Fig.\
\SMhottraj{c}). In both cases, the implied timescales are reduced by
about 40\% with respect to the results at 360~K. Hence, the overall
trends shown by the two contact-based MSMs appear to draw a consistent
picture of the temperature dependence of the system.

%
%
\subsection{Exploring the method space}

To provide an overview of all considered methods and features, we now
introduce three qualitative scoring functions. While the dynamical
quality of an MSM can be readily assessed via the GMRQ
score\cite{McGibbon15} reflecting the implied timescales, it is less
straightforward to characterize the structural quality of a state
partitioning. Here two complementary measures are employed: On the one
hand, we use the Shannon entropy $H = - \sum_j p_j \ln p_j$ of the
state populations $p_j$, which is maximal for equally distributed
$p_j$. Hence, a high value of $H$ means that the majority of states is
well populated, which we favor over partitionings with a few highly
and many lowly populated states. On the other hand, we consider the
Davies-Bouldin index \cite{Davies79} which is a well-established
descriptor of the similarity of clusters (see Methods), and is to be
minimized to ensure structurally distinct states. To obtain
easy-to-compare quantities that all should be maximal, we consider
$\Delta$DBI = DBI$_{\rm max} -$DBI.

\begin{figure}[t!]
    \centering
	\includegraphics[scale=0.9]{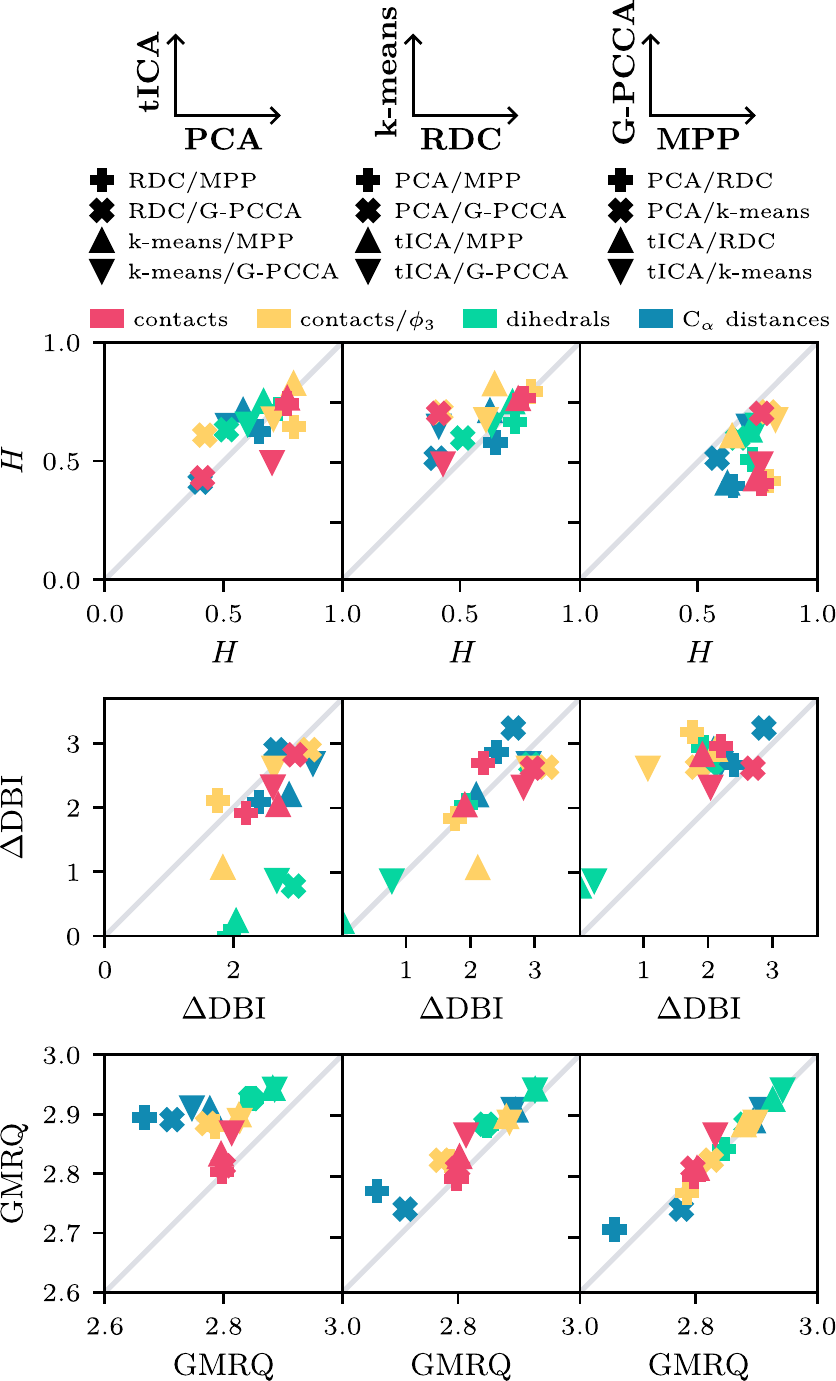}
        \caption{
          Comparison of various methods and features, showing (top)
          the normalized entropy score $H$, (middle) the Davies-Bouldin index
          $\Delta$DBI = DBI$_{\rm max} -$DBI, and (bottom) the
          timescale score GMRQ, which all should be maximized. }
    \label{fig:comparison}
\end{figure}

Comparing various method combinations of the MSM workflow, Fig.\
\ref{fig:comparison} shows these three scoring functions as obtained
for various features, including contacts, contacts/$\phi_3$, backbone
dihedral angles, and selected C$_\alpha$-distances. We begin with the
comparison of PCA and tICA, which as expected shows that tICA clearly
improves the timescale score, in particular for dihedrals. While the
entropy score shows no clear trend, the Davies-Bouldin index reveals
the opposite picture, i.e., PCA generally obtains better scores, and
the tICA results for dihedrals indicates a significant structural
overlap of the resulting states. This well-know effect\cite{Nagel19}
is caused by left to right-handed transitions along the $\phi$
dihedral angles, which represent slow but functionally irrelevant
motions, because the left-handed states are hardly
populated. Nevertheless, these irrelevant motions are favored by tICA
(because they are slow) and therefore lead to structurally nonsensical
states.\cite{Sittel18}

When we compare RDC and $k$-means clustering (middle column of Fig.\
\ref{fig:comparison}), no clear trend is apparent, which is in line
with our findings above. While timescales seem to be sightly better
reproduced by $k$-means, the situation is diverse for the two
structural scores, where sometimes RDC and some other time $k$-means
performs better.
Comparing MPP and G-PCCA (right column of Fig.\ \ref{fig:comparison}),
the resulting timescales are virtually identical.  Since G-PCCA tends
to result in a few well-populated main states, the entropy score
favors MPP, while the Davies-Bouldin index favors G-PCCA because these
few main states are clearly distinct.

%
%
\section{Conclusions}

To validate the contact-based MSM of Ref.\ \citenum{Nagel23}, we have
considered various combinations of features, dimensionality reduction
methods and clustering schemes. Using minimal-distance contacts as
features, we have found that --depending on the specific combination--
we get MSMs with quite similar implied timescales but possibly
different partitionings of the free energy landscape. While G-PCCA was
shown to generally achieve lower spatial resolution than MPP, the
usage of tICA instead of PCA and $k$-means instead of RDC for the most
part resulted in overall comparable state partitionings. In this way, the
overall consistence of the reference model with the alternative method
combinations confirms our initial proposition that the folding of HP35
exhibits a hierarchical structure of the free energy landscape, where
both the native and the unfolded basin split up in various well
populated metastable states, and where a few lowly populated
intermediate states exist.

Hence, our study has demonstrated that for a given MD trajectory
several sets of metastable states with different structural ensembles
may co-exist, and that it is not straightforward to assess which state
partitioning is superior. As a consequence, the above results also
indicate to what extent a reference or benchmark MSM may be defined
for a high-dimensional biomolecular process such as the folding of
HP35. While our reference model is certainly not the only or the best
one, it nonetheless satisfies the above formulated criteria of
achieving sufficient structural detail to resolve the various
metastable states and folding pathways of HP35, and of yielding
sufficiently long implied timescales to represent a correct Markovian
model.

Our study of various methods and features has also revealed which
parts of the MSM workflow matter most for the considered
problem. Notably, we have found that standard methods of
dimensionality reduction (like PCA and tICA), microstate clustering
(such as RDC and $k$-means), and macrostate clustering (such as MPP
and PCCA) typically give reasonable MSMs, if they are applied with
caution (see Methods), and if appropriate features are employed. Using
improper features, the MSM necessarily fails, regardless of what methods
are subsequently used: 'garbage in, garbage out'. For example, we have
shown that Cartesian atom coordinates completely fail to describe the
hierarchical nature of the unfolded region, due to the mixing of
internal and overall motion. On the other hand, we have found that
contact-based MSMs quite reliably produce good and robust timescales
and state partitionings for a variety of methods.

The discussion above points out the importance of the choice and the
pre-selection of suitable features. While the basic idea is to choose
internal coordinates that account for the mechanistic origin of the
studied process, it also means to carefully exclude
irrelevant and deceiving motions from the analysis.  To this end, we
recently proposed the correlation analysis MoSAIC,\cite{Diez22}
which aims to discriminate collective motions underlying
functional dynamics from uncorrelated motion. Applied to dihedral
angles, for example, MoSAIC readily identifies slow uncorrelated
motion along $\phi$ angles,\cite{Nagel23} which would
deceive timescale optimizing approaches such as tICA (see the
discussion of tICA in Fig.\ \ref{fig:comparison}). Moreover, we have
shown that restricting ourselves to selected C$_\alpha$-distances that
approximate interresidue contacts significantly improves the
signal-to-noise ratio, such that details of the free energy landscape
can be resolved. 
While we are confident that the merits of feature selection are just
as important for the modeling of much larger systems than HP35, this
remains to be proven in future work.

%
%
\section{Methods}

{\bf MD details.} This work is based on the
$\SI{300}{\micro\second}$-long MD simulation ($1.5 \times 10^6$ data
points) of the fast folding Lys24Nle/Lys29Nle mutant of HP35 (pdb
2F4K, Ref.\ \citenum{Kubelka06}) at
$T=\SI{360}{\kelvin}$ by Piana et al.,\cite{Piana12} using the Amber
ff99SB*-ILDN force-field \cite{Hornak06, Best09, Lindorff-Larsen10}
and the TIP3P water model.\cite{Jorgensen83}

{\bf Feature selection.}  Features include contacts determined via
minimal distances and via C$_\alpha$-distances, Cartesian atom
coordinates, all C$_\alpha$-distances, backbone dihedral angles, and a
mixture of contacts and angles. Their computation is defined in the
main text. To achieve a pre-selection of features, we employed the
Python package \emph{MoSAIC}\cite{Diez22} (``Molecular Systems Automated
Identification of Cooperativity''). It is based on a community
detection technique called Leiden clustering\cite{Traag19}, which
employs the constant Potts model as objective function.

{\bf Dimensionality reduction.}  We focused on two simple and widely
used linear methods, PCA and tICA, which produce smoothly varying free
energy landscapes that facilitate the subsequent clustering. PCA was
employed as implemented in the open source software \emph{FastPCA},
\cite{Sittel17}. tICA was employed as implemented in PyEmma,
\cite{Scherer15} using $\tau_{\text{lag}}= 10\,$ns and scaling the
eigenvectors by their eigenvalues.\cite{Noe15} Including the first
five components in both cases, subsequently a Gaussian low-pass filter
with a standard deviation of $\sigma=2\,$ns was employed.

{\bf Clustering.}  RDC\cite{Sittel16} was implemented in the open
source software \emph{Clustering}, \cite{Nagel19} using a hypersphere
radius $R = 0.124$ that equals the lumping radius. MPP was implemented
as described in Ref.\ \citenum{Jain12}.  Alternatively, we employed
the mini-batch $k$-means algorithm\cite{Sculley10} in combination with
the $k$-means++ initialization,\cite{Arthur07} as implemented by
Scikit-Learn.\cite{Pedregosa11} We used $k=1000$, a batch size of 5120
frames, 100 random initial configurations, and a minimum (maximum) of
$10^3$ ($10^6$) iteration steps. These relatively high numbers of
configurations and iterations are required to obtain reproducible
results from $k$-means, which can be compared to the deterministic RDC
method.  G-PCCA represents an extension of the well-established PCCA+
approach\cite{Roeblitz13} that is implemented in PyEmma
\cite{Scherer15} and MSMBuilder.\cite{MSMBuilder}

{\bf Analysis.}  All analyses shown in this paper were performed using
our open-source Python package msmhelper.\cite{Nagel23a} To facilitate the
reproduction and analysis of our results for the reference model, we
furthermore provide trajectories of all intermediate steps.

The Davies-Bouldin index for $N$ macrostates is defined by \cite{Davies79}
\begin{align}
	\text{DBI} &= \frac{1}{N} \sum_{i=1}^N \max_{i\neq j}
                     \left( \frac{s_i + s_j}{r_{ij}} \right), \nonumber
\end{align}
where
$s_i = \langle |\boldsymbol{x} - \langle \boldsymbol{x}\rangle_i|
\rangle_i$ is the average distance between each point of a state $i$
and the centroid of that cluster, 
$r_{ij} = |\langle \boldsymbol{x}\rangle_i - \langle
\boldsymbol{x}\rangle_j|$ is the distance between cluster centroids
$i$ and $j$, and $\boldsymbol{x}$ represents the 42 contact distances.

{\bf Computational effort.}  The complete workflow of the reference
MSM took 16 min CPU time on a standard desktop computer, except for
contact definition which additionally took about 2 h. As discussed in
Ref.\ \citenum{Nagel23}, the treatment of proteins with $10^3$
residues is estimated to require in total a few CPU days, which is
negligible compared to the time required for the MD simulations.

%
%
\subsection*{Acknowledgment}

The authors thank Georg Diez, Matthias Post and Steffen Wolf for
helpful comments and discussions, as well as D. E. Shaw Research for
sharing their trajectories of HP35. This work has been supported by
the Deutsche Forschungsgemeinschaft (DFG) within the framework of the
Research Unit FOR 5099 ''Reducing complexity of nonequilibrium
systems'' (project No.~431945604), the High Performance and Cloud
Computing Group at the Zentrum f\"ur Datenverarbeitung of the
University of T\"ubingen and the Rechenzentrum of the University of
Freiburg, the state of Baden-W\"urttemberg through bwHPC and the DFG
through Grant Nos. INST 37/935-1 FUGG (RV bw16I016) and INST 39/963-1
FUGG (RV bw18A004).

\subsection*{Data Availability Statement}
 
The simulation data and all intermediate results for our reference
model, including our software packages \emph{MoSAIC},\cite{Diez22}
\emph{FastPCA},\cite{Sittel17} \emph{Clustering}\cite{Sittel16} and
\emph{msmhelper}\cite{Nagel23a} and detailed descriptions to reproduce
all steps of the analyses, can be downloaded from
https://github.com/moldyn/HP35. In particular, we provide trajectories
($\SI{300}{\micro\second}$ long, $1.5 \times 10^6$ data points) of (1)
the contact distances, (2) the maximal-gap shifted dihedral angles,
(3) the resulting principal components, and (4) the resulting micro-
and macrostates.

\subsection*{Supporting Information}
 
Figure \SMmethods\ shows for all eight considered combinations of
methods (PCA vs.\ tICA, RDC vs.\ $k$-means, and MPP vs.\ G-PCCA) the
MPP dendrogram, the contact characterization of the metastable states,
a Sankey diagram to compare to the reference states, and the implied
timescales. Figure \SMfeatures\ shows these quantities for various
features including Cartesian atom coordinates, selected and all
C$_\alpha$-distances, and contacts/$\phi_3$. Figure \SMhottraj\ shows
results obtained for a folding trajectory at 370~K.

%
%

\begin{mcitethebibliography}{114}
\bibitem[not()]{note3}
{Aiming at a rigorous calculation of reaction coordinates, however, besides
  ($\phi,\psi$) additional internal coordinates were suggested, see, e.g.,
  Refs.\ \citenum{Bolhuis00,Wu22} }\relax
\mciteBstWouldAddEndPuncttrue
\mciteSetBstMidEndSepPunct{\mcitedefaultmidpunct}
{\mcitedefaultendpunct}{\mcitedefaultseppunct}\relax
\EndOfBibitem
\bibitem[not()]{note2}
{In fact, these conditions produced 13 (instead of 12) contacts-based states.
  Since the additional state 13 (on the very left side of the dendrogram) is
  completely unfolded just like state 12, however, we simply merged them into a
  single state when we discuss folding\cite{Nagel23}}\relax
\mciteBstWouldAddEndPuncttrue
\mciteSetBstMidEndSepPunct{\mcitedefaultmidpunct}
{\mcitedefaultendpunct}{\mcitedefaultseppunct}\relax
\EndOfBibitem
\end{mcitethebibliography}

\providecommand{\latin}[1]{#1}
\makeatletter
\providecommand{\doi}
  {\begingroup\let\do\@makeother\dospecials
  \catcode`\{=1 \catcode`\}=2 \doi@aux}
\providecommand{\doi@aux}[1]{\endgroup\texttt{#1}}
\makeatother
\providecommand*\mcitethebibliography{\thebibliography}
\csname @ifundefined\endcsname{endmcitethebibliography}
  {\let\endmcitethebibliography\endthebibliography}{}

\end{document}


\begin{figure}[ht!]
    \centering
    \includegraphics[scale=0.6]{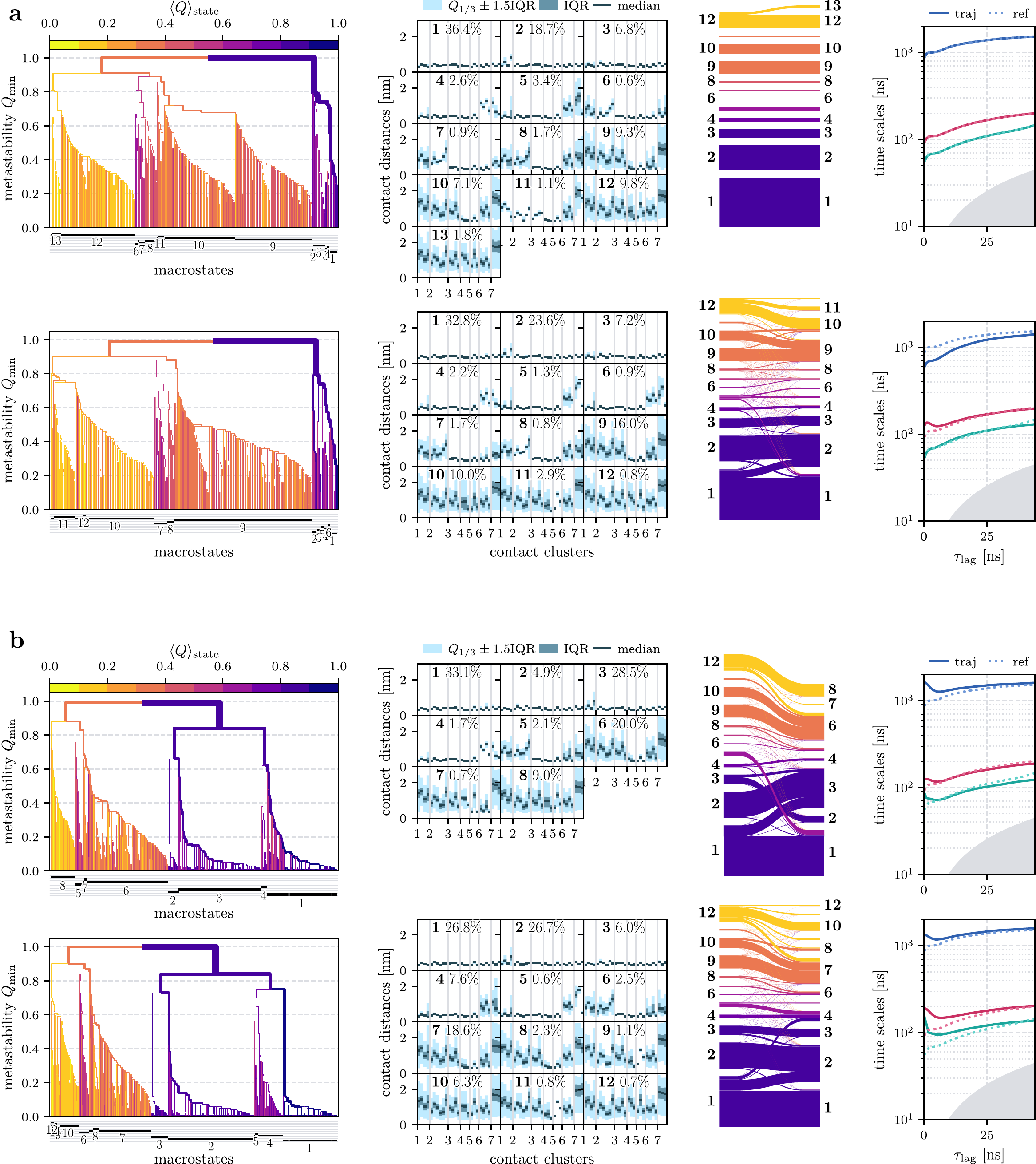}
    \caption{Effects of different method combinations on the
      contact-based Markov state model (MSM) of HP35 established in
      Ref.\ \cite{Nagel23}. Shown are the combinations of (a)
      robust density-based clustering\cite{Sittel16} (RDC) and the
      most probable path algorithm\cite{Jain12} (MPP), (b) $k$-means
      clustering \cite{Sculley10,Arthur07,Pedregosa11} and MPP, (c)
      RDC and generalized Perron cluster cluster
      analysis\cite{Reuter19} (G-PCCA), and (d) $k$-means and G-PCCA.
      Each row shows (1) the MPP dendrogram demonstrating the
      classification of microstates into metastable states (if
      applicable), (2) the contact representation of the resulting
      metastable states, (3) a Sankey diagram contrasting the
      states of the reference model (PCA/RDC/MPP, left) and the states
      from the considered method combination (right), and (4) the
      first three implied timescales drawn as a function of the lag
      time $\tau_{\text{lag}}$. Moreover, each of the sub-figures (a)
      - (d) shows results obtained from (top) principal component
      analysis\cite{Ernst15} (PCA), and (bottom) time-lagged
      independent component analysis \cite{Perez-Hernandez13} (tICA)
      using a lag time of $\tau_\text{lag}=\SI{10}{\nano\second}$.  }
    \label{SI:fig:Methods}
  \end{figure}

\begin{figure}[ht!]
	\ContinuedFloat
	\centering
	\includegraphics[scale=0.6]{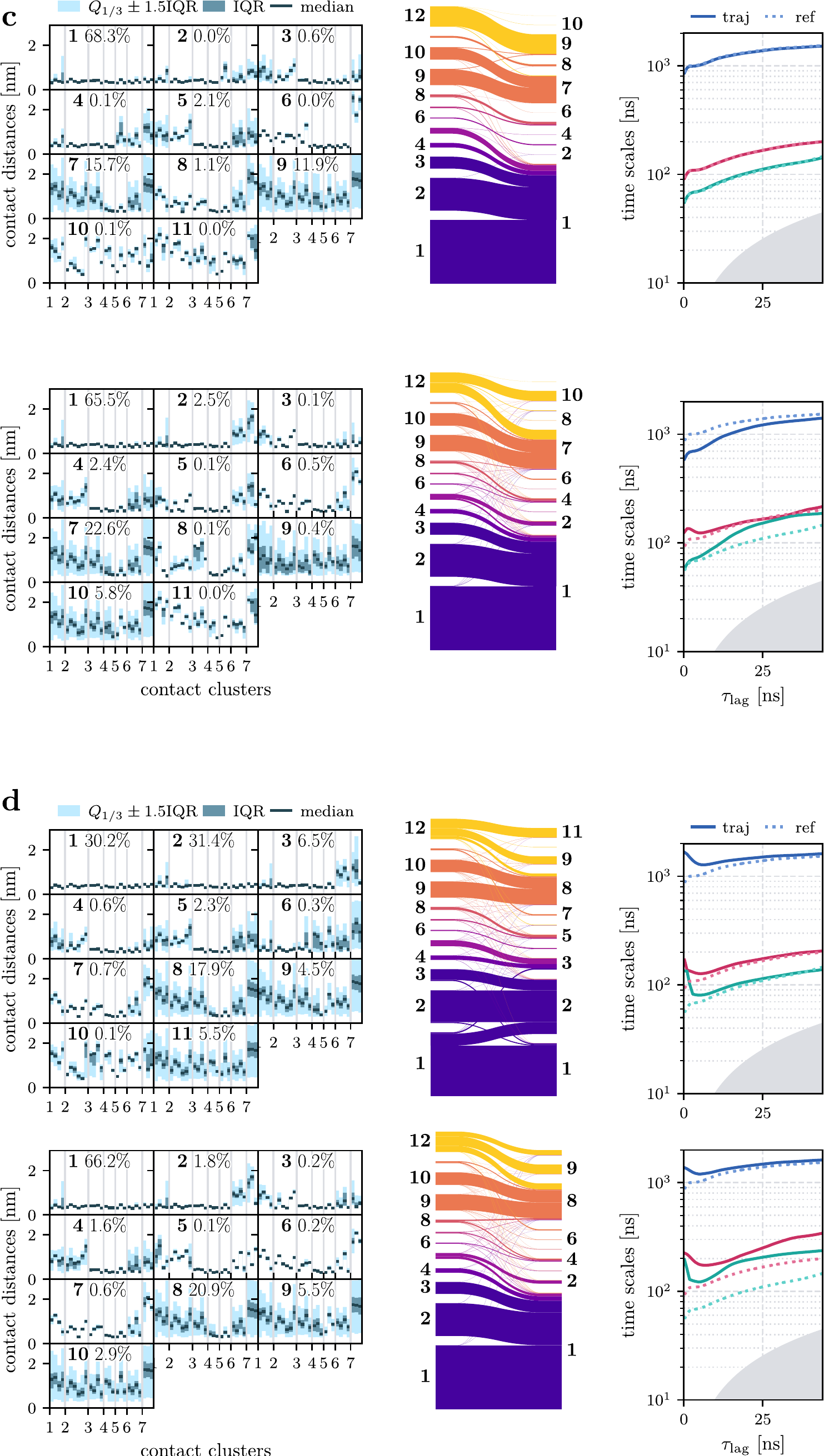}
        \caption{(Continued)}
\end{figure}

\begin{figure}[ht!]
	\centering
	\includegraphics[scale=0.6]{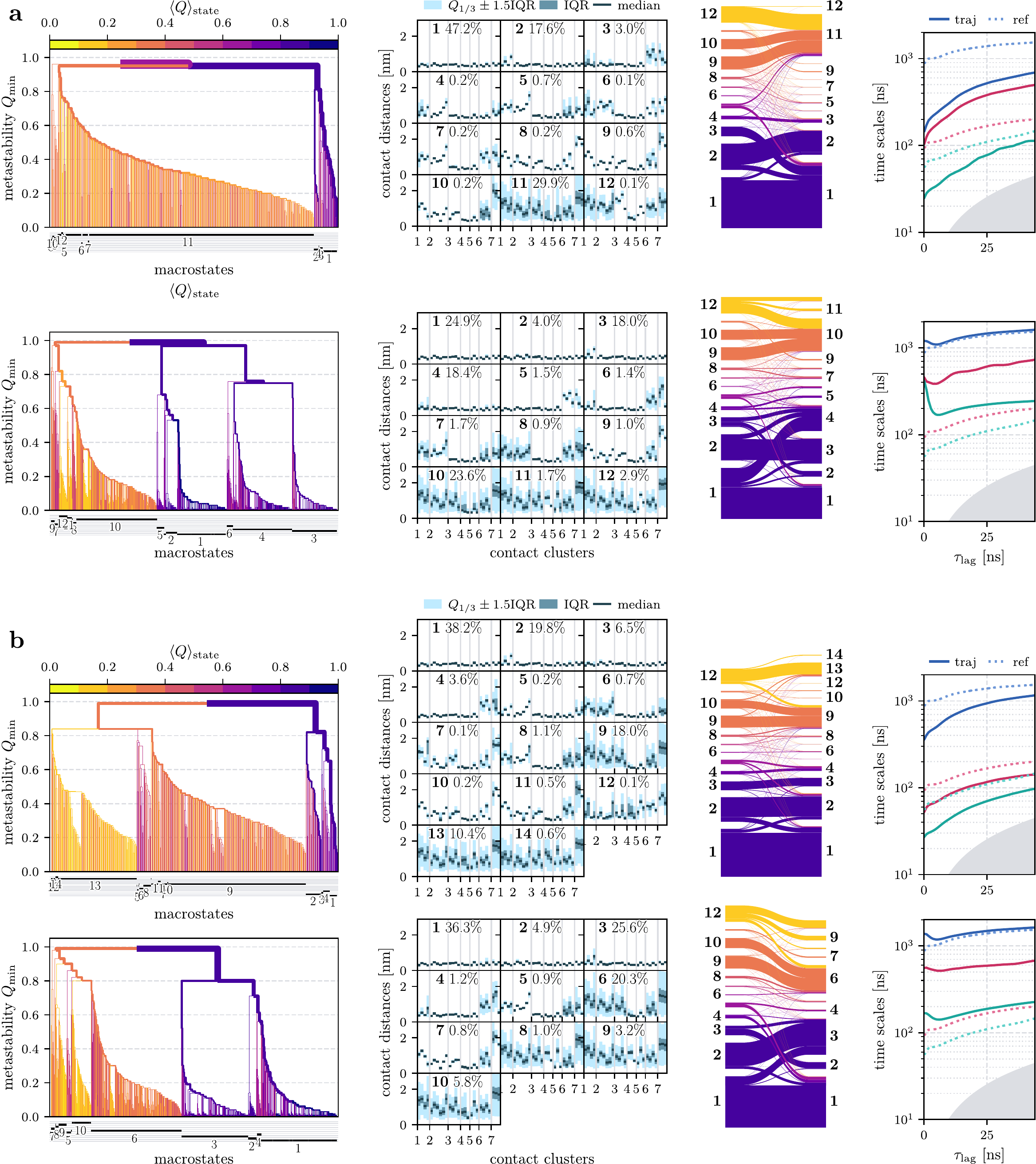}
        \caption{Effects of using different input features on the
          contact-based MSM of HP35, including (a) Cartesian
          coordinates of all $\text{C}_\alpha$-atoms, (b)
          $\text{C}_\alpha$-contacts (i.e., all distances fulfilling
          $P(d_{ij}\le\SI{8}{\angstrom})\ge\SI{30}{\%}$ and
          $|i-j| \ge4$), (c) all $\text{C}_\alpha$-distances, and (d)
          minimal contact distances including the backbone dihedral
          angle $\phi_3$. Each row shows (1) the
      MPP dendrogram demonstrating the classification of microstates
      into metastable states, (2) the contact representation of the
      resulting metastable states, (3) a Sankey diagram
      contrasting the states of the reference model (PCA/RDC/MPP,
      left) and the states from the considered features
      (right), and (4) the first three implied timescales drawn as a
      function of the lag time $\tau_{\text{lag}}$. Moreover, each of
      the sub-figures (a) - (d) shows results obtained from (top)
      PCA, and (bottom) tICA using a lag time of
      $\tau_\text{lag}=\SI{10}{\nano\second}$.
}
	\label{SI:fig:}
\end{figure}

\begin{figure}[ht!]
	\ContinuedFloat
	\centering
	\includegraphics[scale=0.6]{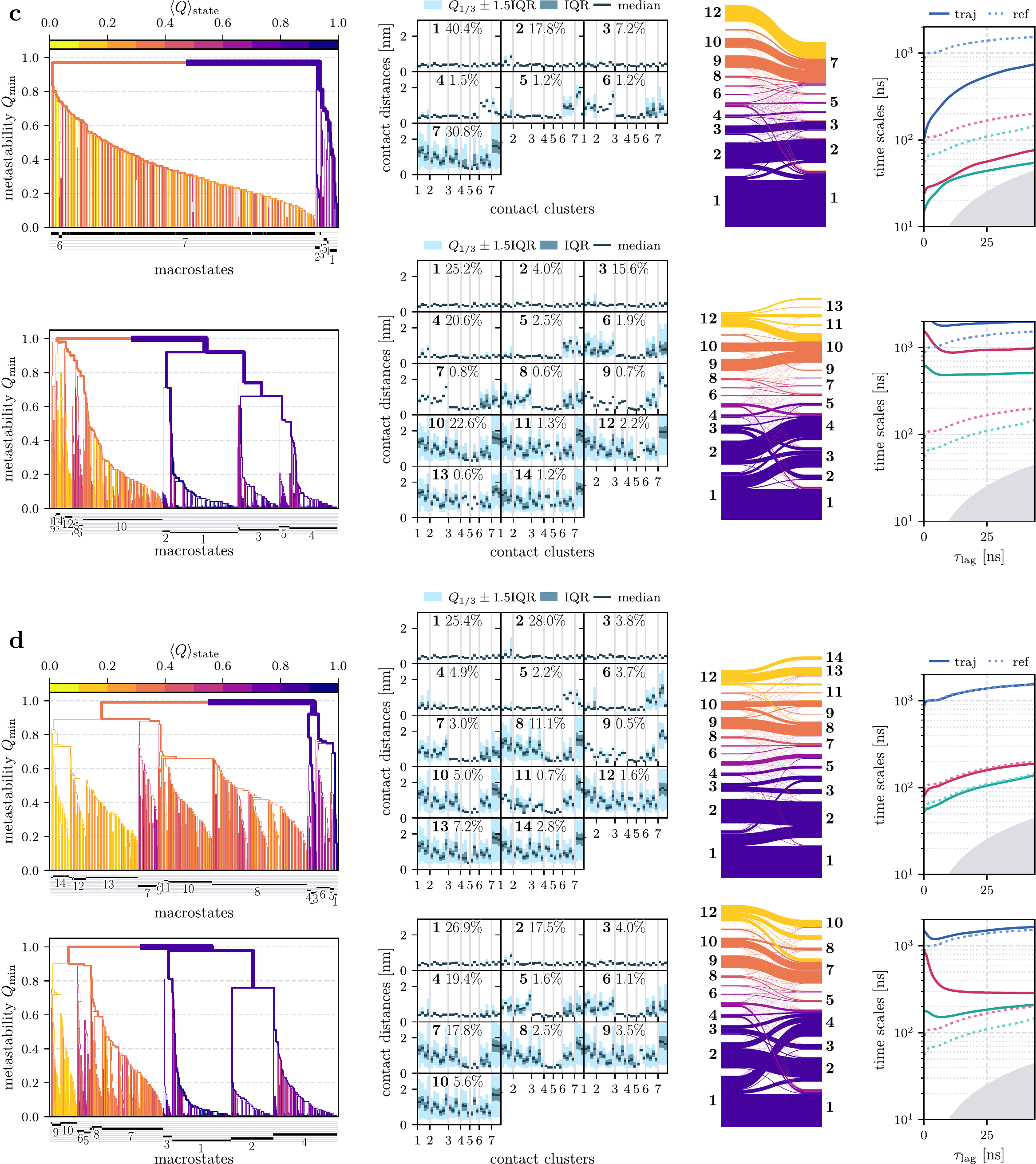}
	    \caption{(Continued)}
\end{figure}

\clearpage
\begin{figure}[ht!]
	\centering
	\includegraphics[scale=0.6]{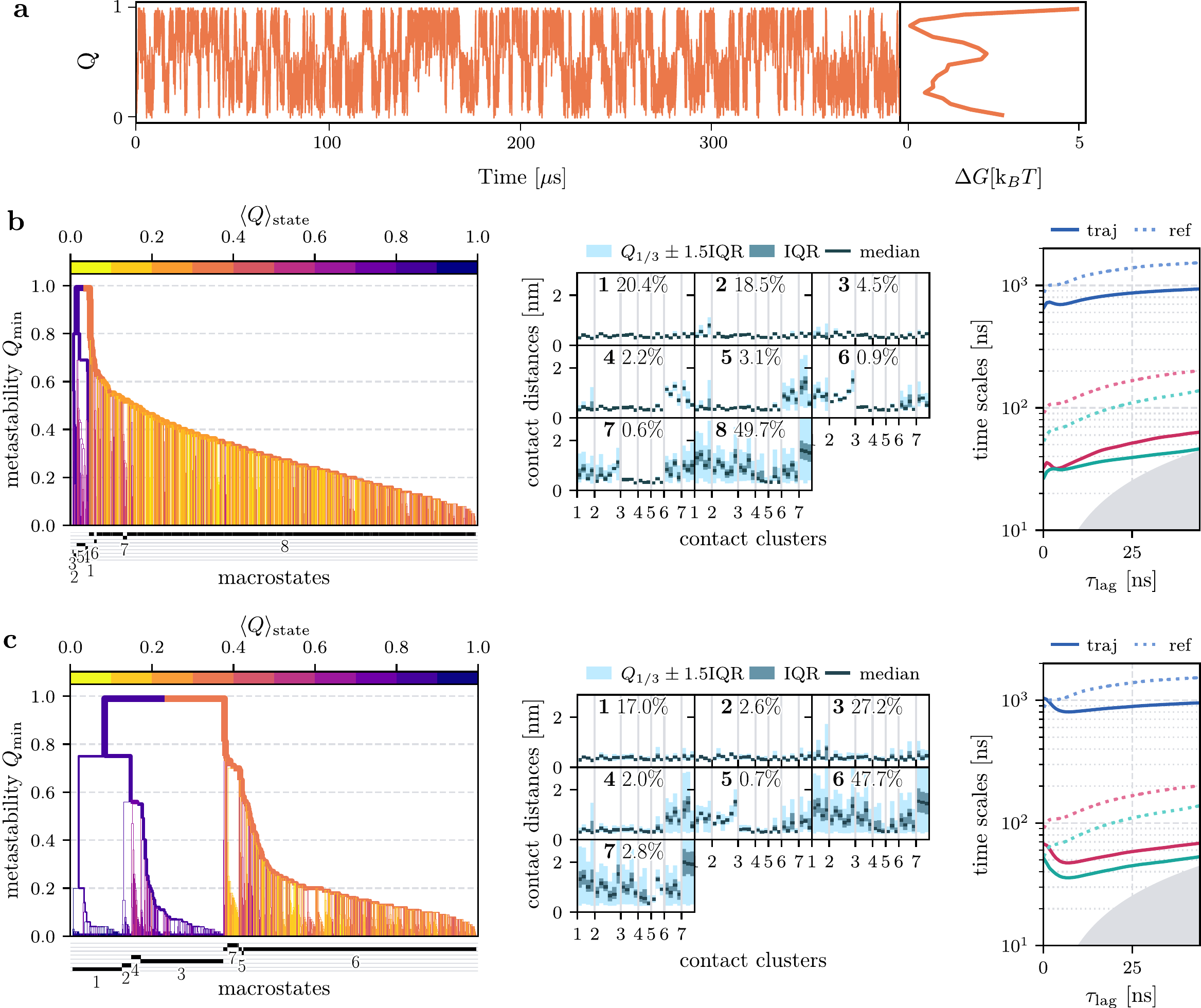}
        \caption{Results obtained for a folding trajectory of HP35 at
          370~K (instead of 360 K) obtained by Piana et
          al.\cite{Piana12} Shown are (a) the time evolution of the
          fraction of native contacts, $Q$, as well as results
          (including MPP dendrogram, contact representation of the
          states, and implied timescales) obtained for (b) our standard
          workflow PCA/RDC/MPP and (c) the the combination
          tICA/$k$-means/MPP. }
\end{figure}

\clearpage
\bibliography{\dir/stock,\dir/md,../new}